\documentclass[epj]{svjour}
\usepackage[dvips]{graphicx}
\begin{document}
\title{Study of light kaonic nuclei with a \\
Chiral SU(3) - based $\bar{K}N$ interaction}
\author{Akinobu Dot$\acute{\rm e}$\inst{1} \and Wolfram Weise\inst{2}
}                     
%
%
\institute{IPNS/KEK, 1-1 Ooho, Tsukuba, Ibaraki, Japan, 305-0801 
\and Physik-Department, Technische Universit\"at M\"unchen, 85747 Garching, Germany}
\date{Received: date / Revised version: date}
%
\abstract{
We have investigated the prototype kaonic nucleus, $ppK^-$, using the 
method of Antisymmetrized Molecular Dynamics (AMD). In the present study 
we use a realistic $NN$ potential with strongly repulsive core, 
and a Chiral SU(3)-based $\bar{K}N$ interaction which is energy-dependent 
and includes both $s$- and $p$-wave interactions. We find that no self-consistent solutions 
exist when the 
range parameter of the $\bar{K}N$ interaction is less than 0.67 fm. 
Due to the strong repulsion of the $NN$ interaction at short distance, 
the two nucleons in $ppK^-$ keep a distance  of about 1.2 fm and 
the total binding energy of $ppK^-$ does not exceed about 50 MeV in the scope of 
our current analysis.
} 
\maketitle

Kaon-nuclear systems, with a $K^-$ tightly bound to a nuclear core, have 
recently received considerable interest.
Previous studies \cite{Akaishi-Yamazaki,AMDK,AMDK2} 
suggested lots of exotic properties of 
light kaonic nuclei, as a consequence of the strongly attractive 
$I=0$ $\bar{K}N$ interaction. 
Several experiments have been performed by various groups 
\cite{Exp:Iwasaki,Exp:Nagae,Exp:Kishimoto,Exp:Herrmann}.
Recently the FINUDA group reported measurements \cite{Exp:Nagae}  
which they interpret as signals 
for the formation of $ppK^-$ clusters, the basic prototypes of kaonic nuclei.
While this interpretation is not without controversy, the quest for deeply bound
kaon-nuclear states has attracted a great amount of activity in the field.

Our investigations are aimed at a detailed understanding of the conditions under which
kaonic nuclei can exist. 
As a first step we explore the $ppK^-$ system,
focusing on the following two points:  
the role of the repulsive core of the $NN$ interaction and the implementation of
a Chiral SU(3)-based $\bar{K}N$ interaction.
If the kaonic nuclei are dense and compact as suggested by previous calculations
\cite{Akaishi-Yamazaki,AMDK,AMDK2}, 
the strong repulsion of the $NN$ interaction at short distance 
is expected to be very important in acting against such compression into very dense
configurations. 
The $\bar{K}N$ interaction is a key ingredient in the study of kaonic nuclei. 
In the present calculations we adopt a theoretically motivated interaction instead of the previous 
phenomenological one \cite{Akaishi-Yamazaki}. The $s$-wave part of this interaction is derived from  
chiral SU(3) theory \cite{Borasoy} and includes the strong energy dependence
resulting from coupled-channel dynamics. An energy dependent $p$-wave interaction 
dominated by the $\Sigma(1385)$ is also incorporated. 

We investigate light kaonic nuclei in a fully microscopic way employing
the variational Antisymmetrized Molecular Dynamics (AMD) method.  
In the AMD framework, 
single nucleon wave functions $|\varphi_i \rangle$ and 
the kaon wave function $|\varphi_{K} \rangle$ 
are constructed as follows;  
\begin{eqnarray} 
&  |\varphi_i \rangle &  = 
\sum_{\alpha=1}^{N_n} C^i_\alpha 
\exp \left[-\nu \left({\boldmath{\mbox{$r$}}}-\frac{{\boldmath{\mbox{$Z$}}}^i_\alpha}{\sqrt{\nu}} 
\right)^2 \right]\; 
|\sigma_i \rangle | \tau^i_\alpha \rangle, \label{eq:N} \\
&  |\varphi_{K} \rangle &  = 
\sum_{\alpha=1}^{N_K} C^{K}_\alpha 
\exp \left[-\nu \left({\boldmath{\mbox{$r$}}}-\frac{{\boldmath{\mbox{$Z$}}}^{K}_\alpha}{\sqrt{\nu}} 
\right)^2 \right]\; 
| \tau^{K}_\alpha \rangle. \label{eq:K}
\end{eqnarray}
The trial wave function of each nucleon (or the kaon) is described as a superposition of 
$N_n$ ($N_K$) Gaussian wave packets whose centers 
$\{ {\boldmath{\mbox{$Z$}}}^i_\alpha\}$ ($\{ {\boldmath{\mbox{$Z$}}}^K_\alpha\}$) are 
different from each other; $\nu$ is a width parameter of the Gaussian 
wave packets; 
$|\sigma_i \rangle$ represents a spin wave function, $|\uparrow\rangle$ 
or $|\downarrow\rangle$; $| \tau^i_\alpha \rangle$ and $| \tau^{K}_\alpha \rangle$
are isospin wave functions and have the following form: 
\begin{eqnarray} 
& | \tau^i_\alpha \rangle&  = 
\left( \frac{1}{2}+\gamma^i_\alpha \right)| {p} \rangle + 
\left( \frac{1}{2}-\gamma^i_\alpha \right)| {n} \rangle, \label{eq:isoN} \\  
& | \tau^{K}_\alpha \rangle&   = 
\left( \frac{1}{2}+\gamma^{K}_\alpha \right) | {\bar{K}^0} \rangle + 
\left( \frac{1}{2}-\gamma^{K}_\alpha \right) |{K^-}  \rangle \label{eq:isoK}
\end{eqnarray} 
where $\{ \gamma^i_\alpha \}$ and $\{ \gamma^K_\alpha \}$ are variational parameters. 

The total wave function is constructed 
by antisymmetrizing a set of nucleon wave functions, Eq. (\ref{eq:N}),
and combining it with the kaon wave function, Eq. (\ref{eq:K}):  
$|\Phi  \rangle = {\cal A}[|\varphi_i(a)\rangle] \otimes 
|\varphi_{K} \rangle$. This wave function is projected onto an eigenstate of 
angular momentum ($J$), isospin ($T$), charge ($C$) and parity ($P$):
$|P_J P_T P_C P_P \; \Phi \rangle$. 
Given the charge-mixed state, Eqs. (\ref{eq:isoN}) and (\ref{eq:isoK}), and 
the charge projection ($P_C$), we can adequately treat 
the $K^- p / \bar{K}^0 n$ mixing which is caused by the $\bar{K}N$ interaction. 
All variational parameters $\{C^i_\alpha, {\bf Z}^i_\alpha, \gamma^i_\alpha; 
C^K_\alpha, {\bf Z}^K_\alpha,$
$\gamma^K_\alpha \}$, which are complex numbers, 
are determined by the frictional cooling equation. 
Details of the AMD method are explained in Ref. \cite{AMDK2}. 

\begin{figure}[t]
\begin{center}
 \includegraphics[width=6cm,angle=-90]{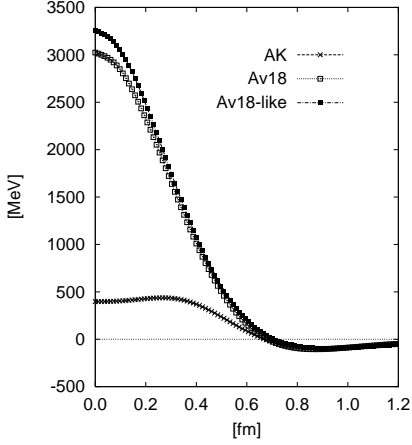}
\end{center}
\caption{Comparison of $NN$ potentials in $^1S_0$ channel. 
``AK'' is the $NN$ potential used in the previous study. 
``Av18'' is the Argonne v18 potential. 
``Av18-like'' is the $NN$ potential used in the present study.
 \label{NNpots}}
\end{figure}

The $NN$ interaction used here is guided by the Argonne v18 (Av18) 
potential \cite{Av18}. 
Fig. \ref{NNpots} shows the $^1S_0$ channel of the potential used in our study (``Av18-like''), 
the original Argonne v18 potential (``Av18''), 
and the potential used in the previous studies (``AK'') \cite{Akaishi-Yamazaki,AMDK,AMDK2}. 
Evidently, the present realistic potential has a strongly repulsive core 
in contrast to the previously used potential. Since this previous potential 
has been constructed using the $G$-matrix method \cite{Akaishi-Yamazaki}, 
the repulsive-core part is smoothed out. 

\begin{figure}[t]
\begin{center}
 \includegraphics[width=6cm,angle=-90]{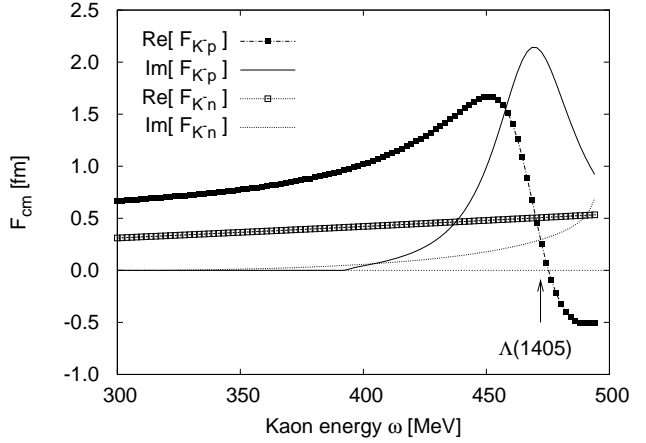}\\
\vspace{0.5cm}
 \includegraphics[width=6cm,angle=-90]{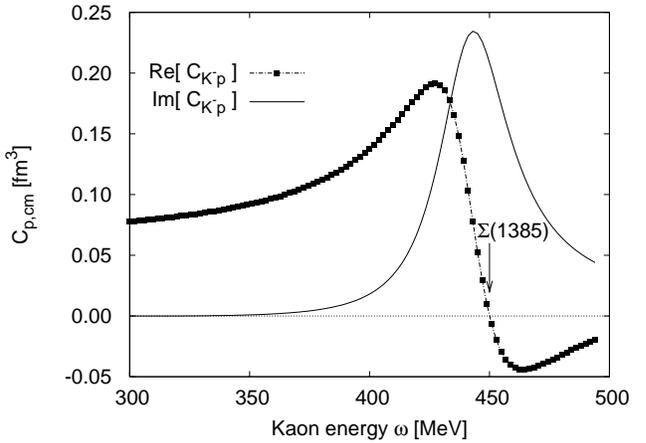}
\end{center}
\caption{The $s$-wave amplitude $F_{KN} (\omega)$ (upper panel) 
and the $p$-wave scattering ``volume" $C_{KN} (\omega)$ (lower panel).
 \label{Scatt}}
\end{figure}

The $s$-wave and $p$-wave $\bar{K}N$ interactions have the following forms: 
\begin{eqnarray}
& & v_{KN,S} ({\boldmath{\mbox{$r$}}}_N-{\boldmath{\mbox{$r$}}}_K, \omega)  = - 2\pi\left({\omega + M_N\over \omega M_N}\right)  F_{KN}(\omega) \nonumber \\
& & \hspace{1.0cm} \times \frac{1}{\pi^{3/2} a_s^3} \exp[-({\boldmath{\mbox{$r$}}}_N-{\boldmath{\mbox{$r$}}}_K)^2/a_s^2], \label{KNs}\\
& & v_{KN,P} ({\boldmath{\mbox{$r$}}}_N-{\boldmath{\mbox{$r$}}}_K, \omega) = - 2\pi\left({\omega + M_N\over \omega M_N}\right) C_{KN}(\omega) \nonumber \\
& & \hspace{1.0cm} \times
\frac{1}{\pi^{3/2} a_p^3} \nabla \exp[-({\boldmath{\mbox{$r$}}}_N-{\boldmath{\mbox{$r$}}}_K)^2/a_p^2]\, \nabla, \label{KNp}
\end{eqnarray}
where $M_N$ is the nucleon mass and $\omega$ is the kaon energy.  The present $\bar{K}N$ potentials have a strong energy dependence. 
The $s$- and $p$-wave potentials have Gaussian shapes with range parameters 
$a_s$ and $a_p$. $F_{KN}(\omega)$ and $C_{KN}(\omega)$ are the $s$-wave $\bar{K}N$ 
scattering amplitude and the energy dependent $p$-wave $\bar{K}N$ scattering ``volume", respectively. The amplitude  $F_{KN}(\omega)$ is derived from Chiral SU(3) theory \cite{Borasoy}. Its energy dependence reflects primarily the strong $\bar{K}N\leftrightarrow\pi\Sigma$ coupled-channels dynamics which drives the $\Lambda(1405)$ resonance. The  $C_{KN}(\omega)$ is an updated version of the parameterization introduced long ago in Ref. \cite{BWT}. It features the dominance of the $\Sigma(1385)$ in $p$-wave $I=1$ channel for which  $C_{K^-n} \simeq 2C_{K^-p}$ holds. Fig. \ref{Scatt} shows $F_{KN}(\omega)$ and $C_{KN}(\omega)$. 

The procedure of the present variational calculation is as follows. 
In a first step we treat the $p$-wave $\bar{K}N$ interaction perturbatively and 
omit the imaginary parts of the $s$- and $p$-wave $\bar{K}N$ interactions. 
The total Hamiltonian $\hat{H}$ is 
\begin{eqnarray}
\hat{H} & = & \hat{H}_0 + {\rm Re} [\hat{V}_{KN, P} (\omega)] \\
\hat{H}_0 & \equiv & \hat{T} + \hat{V}_{NN} + {\rm Re} [\hat{V}_{KN, S} (\omega)] 
+ \hat{V}_{Coul.} - \hat{T}_{CM}. 
\end{eqnarray}
Here, $\hat{T}$, $\hat{V}_{NN}$ and $\hat{V}_{Coul.}$ correspond to 
the total kinetic energy, $NN$ central potential and Coulomb potential, respectively. 
We use all channels of the Av18-like potential as our $\hat{V}_{NN}$, 
although channels other than $^1S_0$ 
turn out not to be important in $ppK^-$.  
$\hat{V}_{KN, S} (\omega)$ and $\hat{V}_{KN, P} (\omega)$ are the $s$- and 
$p$-wave $\bar{K}N$ potentials of Eqs. (\ref{KNs}) and (\ref{KNp}), respectively. 
The center-of-mass motion energy of the 
whole system, $\hat{T}_{CM}$, is subtracted. 
We determine an AMD wave function $|\Phi_0\rangle$ by 
the energy variation for the Hamiltonian $\hat{H}_0$. 
Then we calculate the total energy $E_{tot}$ which is 
the expectation value of the total Hamiltonian $\hat{H}$ 
with this wave function $|\Phi_0\rangle$. 
Secondly, we must take into account the self-consistency of the kaon's energy 
since the present $\bar{K}N$ interaction is strongly energy dependent. 
We obtain the self-consistent solution by the following steps. 
1. We assume a kaon binding energy $B(K)_{assumed}$ as trial input. 
Once the kaon's energy is determined as $\omega = m_K - B(K)_{assumed}$, 
the strength of the $\bar{K}N$ interaction is fixed 
and the Hamiltonian is also determined. 
2. We perform the energy variation and treat 
the $p$-wave $\bar{K}N$ interaction  perturbatively  so as to calculate the total energy. 
We compute separately the kaon binding energy, $B(K)_{obtained}$. 
3. We compare $B(K)_{obtained}$ with $B(K)_{assumed}$. 
If they differ, we return to step 1 
and change the input value of $B(K)_{assumed}$. 
We repeat the cycle 1 to 3 until $B(K)_{obtained}$ coincides 
with $B(K)_{assumed}$. When $B(K)_{obtained} = \\B(K)_{assumed}$, 
a self-consistent solution is found. 

\begin{figure}[t]
\begin{center}
\includegraphics[width=6cm,angle=-90]{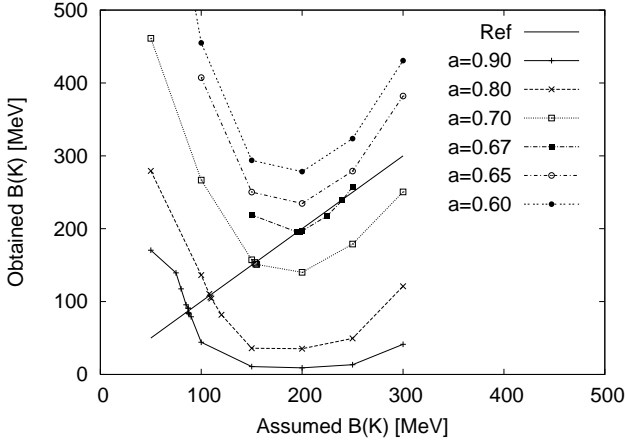}
\end{center}
\caption{Self consistency of the kaon's binding energy $B(K)$. 
The horizontal axis $B(K)_{assumed}$ is the kaon's binding energy 
assumed as input in order to determined the strength of $\bar{K}N$ interaction. 
The vertical axis $B(K)_{obtained}$ is the kaon's binding energy 
calculated with such a $\bar{K}N$ interaction.
``Ref'' is the line $B(K)_{assumed}=B(K)_{obtained}$. The curves correspond 
to results calculated with different range parameters $a$. 
 \label{SelfCons}}
\end{figure}

We comment briefly on the kaon binding energy $B(K)$. 
In kaonic nuclei, $B(K)$ is different from the total energy, $E_{tot}$, 
because the nuclear components rearrange themselves under the influence of 
the strong attraction from the kaon field located at the center of the nucleus. 
In fact, the nuclear part in a kaonic nucleus is in an excited state relative to 
the original nuclear system. 
We define $B(K) = -(E_{tot}-E_{nucl})$ as the kaon binding energy,
where $E_{nucl}$ is the nuclear part of the energy calculated as 
$\langle \Phi_0|\hat{H}_N|\Phi_0\rangle / \langle \Phi_0|\Phi_0\rangle$
with the corresponding part of the Hamiltonian, 
\begin{eqnarray}
\hat{H}_N \equiv \hat{T}_N + \hat{V}_{NN} - \hat{T}_{CM(N)}. 
\end{eqnarray}
Here, $\hat{T}_N$ and $\hat{T}_{CM(N)}$ are the kinetic energy 
and the center-of-mass motion energy of the nucleons only. 

We now show and discuss results for the $ppK^-$ system. 
In the present study, the quantum numbers of 
$ppK^-$ are assumed to be $J^\pi=0^-$ and $T=1/2$, 
where the parity ($\pi$) includes the intrinsic parity 
of the kaon. Single nucleons and the kaon are represented 
by two and five Gaussian wave packets, respectively. 
Namely, we set $N_n=2$ in Eq. (\ref{eq:N}) and $N_K=5$ in Eq. (\ref{eq:K}). 
We use a common range parameter for both $s$-wave and 
$p$-wave $\bar{K}N$ interactions, $a_s=a_p \equiv a$, and 
investigate the systematics with various range parameters $a$.

Fig. \ref{SelfCons} shows the realization of the self-consistency condition for the kaon's 
binding energy $B(K)$, with varying range parameters. 
The self-consistent solution should appear on 
the line ``Ref'' which indicates $B(K)_{assumed}=B(K)_{obtained}$. 
As can be seen, the curves for $a \geq 0.67$ fm cross the line ``Ref''. 
This means there exists a self-consistent solution for these cases. 
However, for $a < 0.67$ fm, no self-consistent solution is found. 

\begin{table}
\caption{Self consistent solution of $ppK^-$ with 
various range parameters. ``$a$'' is the range parameter of 
$\bar{K}N$ interaction in fm. $T$, $V(NN)$, $V(KN,S)$, 
$V(KN,P)$ and $V(Coul.)$ are the total kinetic energy, 
the $NN$ potential energy, the $s$-wave $\bar{K}N$ potential 
energy, the $p$-wave $\bar{K}N$ potential 
energy and  the Coulomb energy, respectively. 
Total E. is the total energy $\langle \hat{H} \rangle$. 
$B(K)$ is the separate binding energy of kaon. 
All energies are given in MeV. 
$R_{pp}$ (in fm) is the mean distance between the two protons. 
}
\label{SumSol}       
\begin{center}
\begin{tabular}{l|rrrr}
\hline\noalign{\smallskip}
          & a=0.67 & a=0.70 & a=0.80 & a=0.90  \\
\noalign{\smallskip}\hline\noalign{\smallskip}
$T$         &  469.8 &  382.2 &  270.4 &  188.8  \\
$V(NN)$     &   15.3 &   16.1 &   14.1 &   12.8  \\
$V(KN,S)$   &$-404.8$&$-340.9$&$-255.0$&$-191.1$ \\
$V(KN,P)$   &$-130.6$&$ -88.8$&$ -49.3$&$ -27.1$ \\
$V(Coul.)$  &$  -2.2$&$  -2.0$&$  -1.7$&$  -1.4$ \\
\noalign{\smallskip}\hline\noalign{\smallskip}
Total E.  &$ -52.5$&$ -33.4$&$ -21.5$&$ -18.0$ \\
$B(K)$    &  195.4 &  153.7 &  110.4 &   84.5  \\
\noalign{\smallskip}\hline\noalign{\smallskip}
$R_{pp}$  &  1.04  &  1.11  &  1.25  &  1.41   \\
\noalign{\smallskip}\hline
\end{tabular}
\end{center}
\end{table}
 
The self-consistent solutions so obtained are summarized in Table \ref{SumSol}. 
As the range parameter decreases, the total binding energy (= $-$ ``Total E.'') 
increases rapidly from 18 MeV to 53 MeV. One observes that the $p$-wave 
$\bar{K}N$ potential contributes significantly to the binding of the kaon.
However, its contribution 
decreases as the range parameter increases. Compared to the 
$s$-wave potential, the $p$-wave part is still sufficiently small to justify a 
perturbative treatment.

In all cases, the average distance between the two protons ($R_{pp}$) is 
larger than 1.0 fm. This is attributed to the strongly repulsive core 
of the $NN$ interaction. If the size of the core of a single nucleon is 
assumed to be roughly 0.5 fm, the present result indicates that the cores 
of the two protons remain well separated in space.

\begin{figure}[t]
\begin{center}
\begin{minipage}[t]{3cm}
\begin{center}
  \includegraphics[width=3cm]{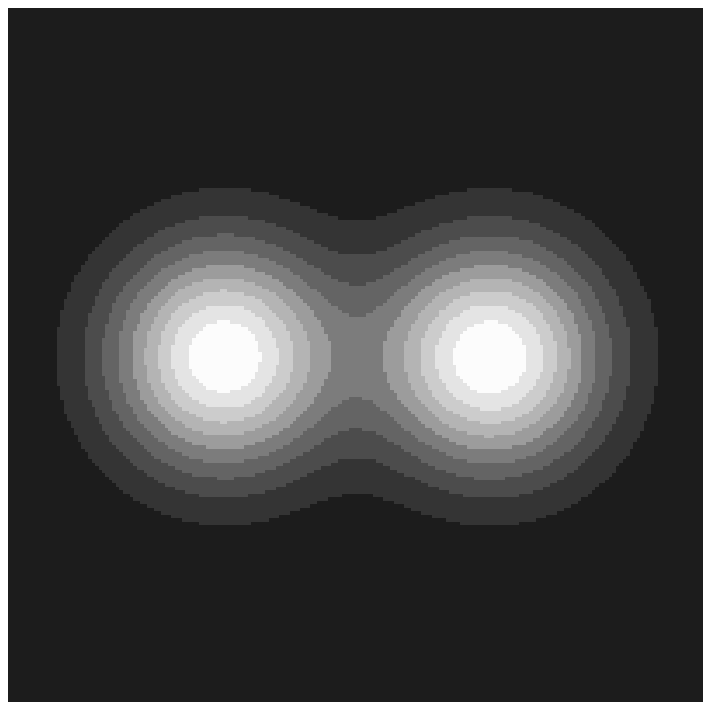}\\
(a) Nucleon
\end{center}
\end{minipage}
\hspace{0.5cm}
\begin{minipage}[t]{3cm}
\begin{center}
  \includegraphics[width=3cm]{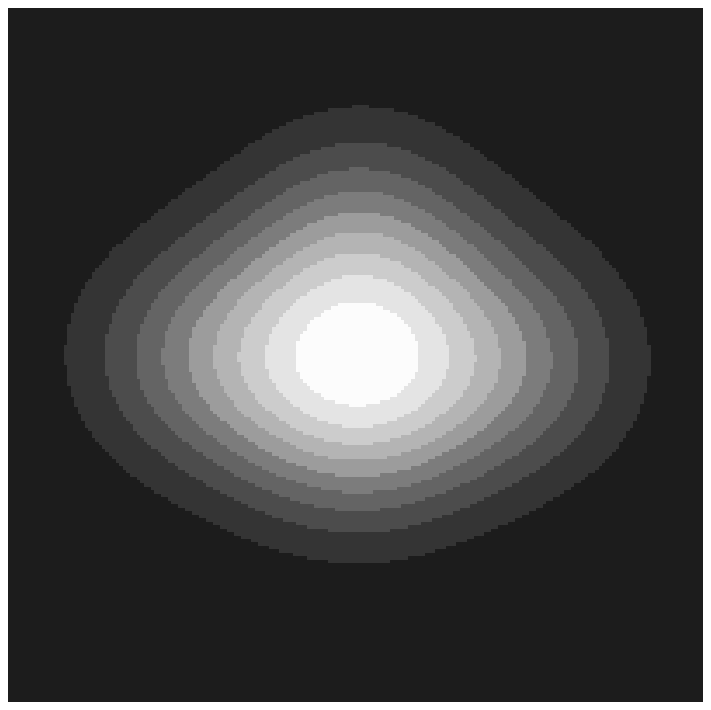}\\
(b) Kaon
\end{center}
\end{minipage}
\end{center}
\caption{Density contour of $ppK^-$ calculated with the 
range parameter a=0.90 fm. The left (right) panel shows 
the distribution of the nucleons ( the kaon). The size scale is $3 \times 3$ fm$^2$.
 \label{Dens_ppK}}
\end{figure}

Fig. \ref{Dens_ppK} depicts the density distribution of the nucleons 
and the kaon in $ppK^-$ calculated with the range parameter $a=0.9$ fm. 
The two protons keep a distance and 
the kaon is centered between them. In cases using other range parameters, 
the configuration of nucleons and kaon is essentially the same as 
that shown in Fig. \ref{Dens_ppK}. 

In Table \ref{OnlyS}, we show a self-consistent 
solution obtained by using 
only the $s$-wave $\bar{K}N$ interaction (``Only $s$-wave''). 
Here the range parameter is 0.70 fm. 
We performed the same calculation as mentioned before, 
switching off the $p$-wave interaction. In this case, 
the total energy is similar to that obtained with 
the $p$-wave interaction (``With $p$-wave''). However, 
the decomposition of this total energy into its components is quite different. 
In the case ``With $p$-wave'', the $ppK^-$ tends to increase its binding energy 
by the additional attraction of the $p$-wave potential, but 
simultaneously reduces binding due to the large kinetic energy, 
compared to the case with ``Only $s$-wave''.  
The proton distance ($R_{pp}$) of ``With $p$-wave'' is 
smaller than that of ``Only $s$-wave''. This is found also 
for other range parameters. The $p$-wave 
interaction prefers a smaller system in order to utilize larger
gradients of the density distributions.

In summary, as a result of the present study of $ppK^-$ with the AMD method  
using the Av18-like $NN$ potential and the Chiral SU(3)-based 
$\bar{K}N$ potential, 
no self-consistent solutions are found when the range parameter 
of the $\bar{K}N$ interaction is less than 0.67 fm. The maximum 
total binding energy of $ppK^-$ is 53 MeV. The configuration of the 
$ppK^-$ cluster is such that the 
two protons keep a distance larger than 1 fm due to the strong repulsion 
of the short-range $NN$ interaction. The kaon field occupies the space between 
them. 

The present results should still be considered preliminary.
The treatment of the short-range 
correlation is thought to be insufficient because the AMD 
variational wave function is based on the independent-particle 
picture. In the next forthcoming step a two-nucleon correlation
function will be introduced in the AMD trial wave function. 
Since the binding mechanism of $ppK^-$ is  
dominated by the strong attraction of the $\bar{K}N$ 
interaction, it is nonetheless expected that the essential points of the 
present study will not be modified substantially once the short-range correlations
are adequately dealt with.

\begin{table}
\caption{Only $s$-wave interaction vs 
including $p$-wave interaction.
``With $p$-wave'' (``Only $s$-wave'') is the result calculated 
with (without) the $p$-wave interaction for $a=0.70$ fm.} 
\label{OnlyS}       
\begin{center}
\begin{tabular}{l|rr}
\hline\noalign{\smallskip}
          & With $p$-wave & Only $s$-wave  \\
\noalign{\smallskip}\hline\noalign{\smallskip}
$T$         &  382.2 &  195.3   \\
$V(NN)$     &   16.1 &   18.9   \\
$V(KN,S)$   &$-340.9$&$-243.1$  \\
$V(KN,P)$   &$- 88.8$&$   0.0$  \\
$V(Coul.)$  &$  -2.0$&$  -1.4$  \\
\noalign{\smallskip}\hline\noalign{\smallskip}
Total E.  &$ -33.4$&$ -30.3$    \\
$B(K)$    &  153.7 &  106.2     \\
\noalign{\smallskip}\hline\noalign{\smallskip}
$R_{pp}$  &  1.11  &  1.36      \\
\noalign{\smallskip}\hline
\end{tabular}
\end{center}
\end{table}

A further important step will be the calculation of the decay width. This involves
not only the mesonic mode ($\bar{K}N \rightarrow \pi Y$) 
but also the two-nucleon absorption process ($\bar{K}NN \rightarrow YN$). 
A reliable estimate of the width of antikaon-nuclear states is crucial in order
to assess the experimental observability of kaonic nuclei.
\vspace{0.5cm}   

We thank Avraham Gal for fruitful and stimulating discussions during his visit in Munich. 
This work is supported in part by BMBF and GSI.

%

\begin{thebibliography}{}
%
%

\bibitem{Akaishi-Yamazaki} Y. Akaishi and T. Yamazaki, 
Phys. Rev. {\bf C65}, (2002) 044005.

\bibitem{AMDK} A. Dot$\acute{\rm e}$, H. Horiuchi, Y. Akaishi 
and T. Yamazaki, Phys. Lett. {\bf B590}, (2004) 51.

\bibitem{AMDK2} A. Dot$\acute{\rm e}$, H. Horiuchi, Y. Akaishi 
and T. Yamazaki, Phys. Rev. {\bf C70}, (2004) 044313. \label{AMDK2}

\bibitem{Exp:Iwasaki} T. Suzuki, Phys. Lett. {\bf B 597}, (2004) 263; 
arXiv:nucl-ex/0310018.  

\bibitem{Exp:Nagae} M. Agnello {\it et al.}, Phys. Rev. Lett. 
{\bf 94}, (2005) 212303. 

\bibitem{Exp:Kishimoto} T. Kishimoto {\it et al.}, Prog. Theor. Phys. 
Suppl. {\bf 149}, (2003) 264; Nucl. Phys. {\bf A754}, (2005) 383. 

\bibitem{Exp:Herrmann} N. Herrmann, Proceedings of ``International 
Conference on Exotic Atom and Related Topics (EXA05)'' published 
by Austrian Academy of Science Press, pp. 73-81. 

\bibitem{Borasoy} B. Borasoy, R. Ni{\ss}ler and W. Weise, 
Eur. Phys. J. A {\bf 25}, (2005) 79. 

\bibitem{Av18} R. B. Wiringa, V. G. J. Stoks and R. Schiavilla, 
Phys. Rev. {\bf C51}, (1995) 38.

\bibitem{BWT}
R. Brockmann, W. Weise and L. Tauscher, Nucl. Phys. {\bf A308}, (1978) 365;
W. Weise, these Proceedings.



\end{thebibliography}
%

\end{document}